# Magnonic spin-transfer torque MRAM with low power, high speed, and error-free switching

Niladri N. Mojumder[1,2], David W. Abraham[1], Kaushik Roy[2], and D. C. Worledge[1]

[1] IBM T. J. Watson Research Center, Yorktown Heights, NY 10598, USA

[2] Department of Electrical and Computer Engineering, Purdue University, West Lafayette, Indiana 47907, USA

**Abstract:** A new class of spin-transfer torque magnetic random access memory (STT-MRAM) is discussed, in which writing is achieved using thermally initiated magnonic current pulses as an alternative to conventional electric current pulses. The magnonic pulses are used to destabilize the magnetic free layer from its initial direction, and are followed immediately by a bipolar electric current exerting conventional spin-transfer torque on the free layer. The combination of thermal and electric currents greatly reduces switching errors, and simultaneously reduces the electric switching current density by more than an order of magnitude as compared to conventional STT-MRAM. The energy efficiency of several possible electro-thermal circuit designs have been analyzed numerically. As compared to STT-MRAM with perpendicular magnetic anisotropy, magnonic STT-MRAM reduces the overall switching energy by almost 80%. Furthermore, the lower electric current density allows the use of thicker tunnel barriers, which should result in higher tunneling magneto-resistance and improved tunnel barrier reliability. The combination of lower power, improved reliability, higher integration density, and larger read margin make magnonic STT-MRAM a promising choice for future non-volatile storage.

## I. Introduction:

Over the last few decades, the trend towards increased adoption of embedded memory, driven by the need to increase the bandwidth of high performance processors, has prompted research in novel memory technologies. One of these, spin-transfer torque magnetic random access memory (STT-MRAM), has stimulated significant interest due to its unique combination of properties, including data non-volatility, unlimited endurance, low power, high performance, and high integration densities, unlike magnetic field driven MRAM [1-4].

An electrical current passing through a pair of ferromagnetic electrodes separated by a metallic spacer or tunnel barrier exerts pseudo-torque on the magnetic moment of the individual electrodes [5, 6]. The magnitude of this spin-transfer torque is proportional to the electrical current [5-8]. Switching using this torque typically requires a pulse length of at least 5-10 ns, unless the device is driven with currents significantly higher than the switching threshold [9, 10]. For applications to MRAM, such high currents are incompatible with the requirement to avoid breakdown of the tunnel barrier during the desired 10 year lifetime, typically achieved by limiting the write voltage across the barrier to around 400 mV and using a tunnel barrier with resistance-area (RA) product in the range of 5 – 10 $\Omega$–$\mu m^2$ [11]. Lower RA barriers are more susceptible to breakdown and have lower tunneling magneto-resistance (TMR), which reduces read margin. Lowering the switching threshold by, for example, thinning the magnetic storage layer, would reduce the activation energy, causing poor retention due to thermally activated switching. Hence, it is difficult to achieve fast switching of the free layer using STT without increasing errors due to tunnel barrier breakdown, read, and thermal instabilities.

Recently, Slonczewski proposed initiating spin-transfer torque by applying a heat flux through the free layer and an insulating reference ferrite [12]. This thermagnonic torque is generated by creation of magnons in the ferrite and subsequent conversion to electron spin current through the conducting ferro-magnet. Depending upon the direction of heat flow, this spin-transfer torque tends to align or anti-align the free layer magnetic moment with the ferrite magnetic moment. With proper suppression of heat energy carried by phonons inside the ferrite [13, 14], the proposed thermagnonic spin-transfer torque is predicted to have a quantum yield almost two orders of magnitude higher than achievable using conventional electric current through the magnetic tunnel junction (MTJ) [12]. This potentially allows the development of new spintronic devices with low power and high speed operation. However, since the sign of the thermagnonic spin-transfer torque depends only on the direction of heat current, but not on the direction of the electric current used to generate it, the proposed thermagnonic spin-transfer is not directly applicable to MRAM, where bi-directional switching of the free layer is required.

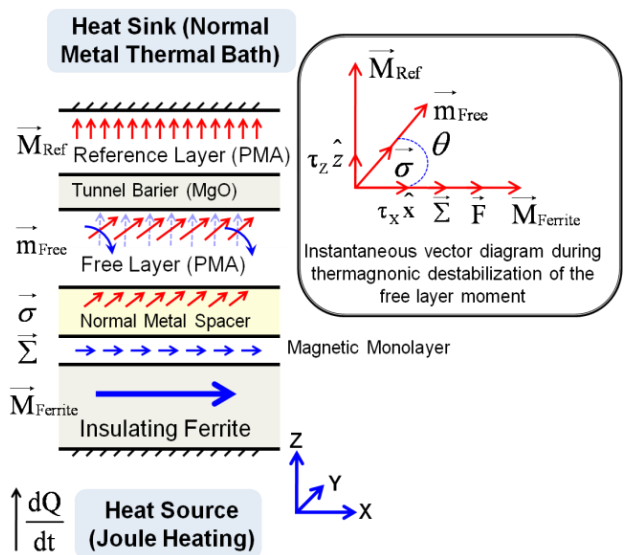

**Fig. 1** Schematic of the proposed magnetic tunnel junction

In this paper, we propose an alternative MRAM using thermagnonic spin-transfer torque that offers reduced switching energy, faster speed, and reliable error-free bi-directional switching of the free layer magnetization. The proposed design shown in Fig. 1 allows for bi-directional switching by making several design alterations to the original multi-layer structure as proposed by Slonczewski [12]. Information is stored in a magnetic free layer with perpendicular magnetic anisotropy (PMA). An insulating ferrite film with uniaxial magnetic anisotropy orthogonal to that of the adjacent free layer is used to destabilize the initial free layer moment during write. During destabilization, the free layer magnetization is rotated into the film plane by thermagnonic spin-transfer torque. Destabilization is achieved by applying an electric pulse to the Joule heater, which creates a heat current that passes through the ferrite, which in turn generates a population of magnons (spin-waves). Conversion from magnons to electrical spins occurs at the ferrite/metal-spacer interface, producing a spin current which tends to drive the free layer magnetic moment to a point of metastability (collinear to the ferrite, and orthogonal to the free layer magnetic anisotropy). Note that this destabilization occurs for free layer magnetization initially in either perpendicular orientation, while using unidirectional heat flow through the ferrite film. Next, the heat current is removed and is followed immediately by a small electric current through the magnetic tri-layer (free layer, tunnel barrier, and reference layer) which flips the free layer moment towards one of the intended stable orientations (parallel or anti-parallel to the reference layer). This electric current requires a third terminal connected to the normal metal spacer.

The major contributions of the present communication are:

- Design proposal of a three-terminal MTJ using both thermagnonic and electrical spin-transfer torque for high speed, low power, and almost error-free switching
- Calculation of the effect of thermagnonic spin momentum transfer on the magnetization switching dynamics of the free layer
- Optimization of the electro-thermal circuit of the proposed MTJ for maximum thermagnonic spin-torque efficiency for high switching speed
- Read optimization of proposed thermagnonic MRAM to achieve reduced read disturb, improved read margin, and enhanced tunnel barrier reliability
- Comparison of the proposed thermagnonic MRAM with conventional electrical STT-MRAM

The rest of the paper is organized as follows. In Section II, we review Slonczewski's work on the physics of thermagnonic spin momentum transfer. The impact of thermagnonic spin-transfer torque on the magnetization switching dynamics of the free layer is analyzed in Section III. In Section IV, we propose and compare two CMOS-compatible designs for the proposed electro-thermal device. Read optimization of the proposed MRAM is carried out in section V. Finally, section VI concludes the paper.

**II. Thermal Initiation of Magnonic Spin-Transfer: Physics and Principals of Device Design**

Fig. 1 shows the schematic of one of the possible device configurations to utilize thermagnonic spin-momentum transfer for bi-directional magnetic switching. The uniaxial magnetic anisotropy of the insulating ferrite is oriented in-plane, along the x axis. The reference and free magnetic layers have perpendicular magnetic anisotropy (PMA) and are oriented perpendicular to the plane, along the z axis. In the following, ***bold-italic*** fonts are used to represent vectors. In the proposed device structure shown in Fig. 1, we estimate thermally initiated magnonic spin-transfer torque by considering the exchange coupling between the ferrite moment (***M**$_{Ferrite}$*) and conductive s-electrons in the adjacent metal spacer through an interfacial atomic monolayer with paramagnetic 3d-electron-spin moments ($\Sigma$) [12, 15]. In Slonczewski's model, ***σ*** and ***F*** are respectively the average thermal moment of the s-electron-spins per unit area of the metal spacer and the effective molecular field at the ferrite-metal interface. ***M**$_{Ref}$* and ***m**$_{Free}$* are respectively the magnetic moments of the ferromagnetic reference and free layers, separated by a nanometer thick tunnel barrier. At any given time, t, during thermagnonic spin-momentum transfer from the ferrite to the free layer, the relative angle between ***M**$_{Ferrite}$* and ***m**$_{Free}$* is defined to be θ(t).

Following Slonczewski, the transient behavior of ***σ*** can be captured by considering its coupling to the monolayer moment $\Sigma$ and solving the Bloch equation as written below in vector form [12, 16]:

$$\frac{d\vec{\sigma}}{dt} + \lambda \vec{\Sigma} \times \vec{\sigma} = v_{ds}\delta\vec{\Sigma} - v_s\vec{\sigma} \quad \text{with} \quad (1.a)$$

$$v_{ds} = k_B T \left(\frac{\pi}{\hbar}\right)(J_{sd}\rho)^2 \quad (1.b)$$

In equation 1, $\nu_{ds}$ represents the d-orbital to s-orbital spin relaxation rate, $J_{sd}$ is the on-site s-d exchange coupling and $\rho$ is the s-electron density per atom. The parameters $\lambda$ and $\nu_s$ in equation (1.a) govern the precession of the s-electron moment, $\sigma$, and the effective spin relaxation, respectively. The multi-layer structure shown in Fig. 1 starts out in thermal equilibrium, when no heat current flows through the insulating ferrite and the metal spacer (dQ/dt=0). A temperature differential ($\delta T$) between the ferrite spins (including the magnetic monolayer with 3d-spins) and the adjacent metal creates an interfacial heat current, which drives the system out of equilibrium. A steady flow of heat current yields a non-vanishing value of $d\sigma/dt$, defined as the thermagnonic spin-transfer torque per unit area. The thermagnonic torque can be expressed as $\tau = \tau_x \mathbf{x} + \tau_z \mathbf{z}$ with two vector components:

(1) An in-plane component ($\tau_x$) trying to align or anti-align the free layer moment ($m_{Free}$) with $M_{Ferrite}$ dependent on the direction of heat flow.
(2) A torque component ($\tau_z$) perpendicular to the individual film interfaces, along the z direction.

*A. Estimation of the In-Plane Thermagnonic Torque ($\tau_x$):*

As discussed in [12], the quantum yield of the thermally initiated magnonic spin-transfer torque depends strongly on the relative suppression of heat flow by phonons compared to magnons inside the ferrite. A larger magnonic contribution to the total heat flow across the ferrite-metal interface is desired for higher torque magnitude, for a given thermal energy input. In practice, the thermal conductance due to interfacial phonon scattering [17], and ferrite-to-monolayer and monolayer-to-metal heat transfer efficiencies jointly determine the effective spin-momentum transfer due to ferrite magnons [18]. The magnitude of the in-plane thermagnonic spin-transfer torque can be estimated as [12]:

$$|\tau_x| = \frac{\pi . S.(S+1).N_d.(J_{sd}\rho)^2.F(T)}{3.\hbar.T}|\delta T| \quad (2)$$

Where, $S$= Spin quantum number of the paramagnetic metallic atoms in the magnetic monolayer,

$N_d$= Number of magnetic ions or atoms per unit area of the magnetic monolayer,

$F(T)$= Molecular-field exchange splitting of the magnetic ions sitting at the ferrite-metal interface at temperature T,

$\delta T$= Temperature differential across the ferrite-metal interface, which is sign dependent on the direction of heat flow (positive for heat current flowing from the metal spacer to the ferrite).

Precise estimation of the quantum yield associated with thermagnonic spin-transfer torque is difficult due to experimental uncertainty in the monolayer exchange splitting [$F(T)$] for different ferrite films [19-21]. We conservatively chose the ferrite film with one of the lowest reported monolayer exchange splitting for our designs and simulations; at T=300K, $MnFe_2O_4$ exhibits F = 10meV.

*B. Estimation of the Perpendicular Torque ($\tau_z$):*

Slonczewski estimates the perpendicular component of the thermagnonic torque as:

$$|\tau_z| = \lambda \sum \sigma \sin(\theta) \quad (3.a)$$

$$\sigma = \frac{\tau_x \cos(\theta)}{\nu_s} \quad (3.b)$$

With a thin tunnel barrier ($t_{ox} \sim 1$ nm), the perpendicular component of the spin-wave disperses in the thermal sink on top of the film stack in Fig. 1 [12]. In a quasi-ballistic picture, $\nu_s$ becomes sufficiently large to diminish the effect of $\tau_z$ on the free layer moment $m_{Free}$ [12]. With no significant impact on the free layer moment during switching, we neglect the contribution of the perpendicular torque for all our simulations in the present work.

It is worthwhile to note that the signs of the in-plane ($\tau_x$) and perpendicular ($\tau_z$) components of the thermagnonic torque depend on the direction of heat flow, but not on the direction of any electric current. In the model proposed in Fig. 1, a positive $\delta T$ (= $T_{Ferrite}$ - $T_{MS}$) leads to a negative $\tau_{x0}$, that tends to anti-align free layer moment with $M_{Ferrite}$. For negative $\delta T$, a positive value of $\tau_{x0}$ aligns $m_{Free}$ with $M_{Ferrite}$. In our design, however, the ferrite uniaxial anisotropy is orthogonal to that of the free layer. Therefore the magnitude of $\delta T$ is important in destabilizing the initial free layer moment, while its sign (i.e. the heat flow direction) is not.

**III. Impact of Thermagnonic Spin-Transfer on Free Layer Moment: Simulation and Device Design**

The impact of thermagnonic spin-transfer torque on magnetic precession of the free layer is analyzed by solving the Landau-Lifshitz-Gilbert-Slonczewski (LLGS) equation [5, 12, 22] with an additional in-plane magnonic torque component given in Eq. 2. We perturb the free layer moment in our proposed device (Fig. 1) with $\delta T$ as input, instead of the electric current used in conventional STT-MRAM. For all subsequent single-domain magnetic simulations, we assume S=2.5, $N_d$=4x10$^{18}$cm$^{-3}$, $J_{sd}$=-500meV, $\rho$=150meV, F=10meV, T=300K, and a lithographic line-width of 30nm [12]. The LLGS equation, capturing the time dependent precessional evolution of $m_{Free}$ due to thermagnonic excitation can be written as:

$$\frac{d\vec{m}_{Free}}{dt} = -|\gamma|(\vec{m}_{Free} \times \vec{H}_{eff}) + \alpha(\vec{m}_{Free} \times \frac{d\vec{m}_{Free}}{dt}) + |\tau_{x0}|[-\alpha(\vec{m}_{Free} \times \vec{M}_{Ferrite}) + (\vec{m}_{Free} \times \vec{m}_{Free} \times \vec{M}_{Ferrite})] \quad (4.a)$$

$$\vec{H}_{eff} = H_{x,External} . \hat{x} + H_{y,External} . \hat{y} + [H_{z,External} + H_{Ku2} - H_{Ks}] . \hat{z} \quad (4.b)$$

$$H_{Ku2} = 2\frac{K_{u2}}{M_S} \quad (4.c)$$

$$H_{Ks} = 4\pi M_S \quad (4.d)$$

plane anisotropy fields respectively. $H_{x,External}$, $H_{y,External}$ and $H_{z,External}$ are the three spatial components of the external magnetic field applied to the free layer. We assume the perpendicular anisotropy energy density ($K_{u2}$) and saturation magnetization ($M_S$) to be $4.7 \times 10^6 erg/cm^3$ and $850 emu/cm^3$ respectively, unless specified otherwise.

In Fig. 2 (a-d), we analyze the angular evolution of three spatial components of $m_{Free}$ with two different amplitudes of steady state thermal excitation ($\delta T$), turned on abruptly at t=0. An insufficient thermal excitation ($\delta T=7K$) initiates angular precession ($<<\pi/2$) of the free layer moment around its easy axis, but does not align the free and ferrite moments (*partial destabilization*). However, once the thermal excitation exceeds a certain critical value ($\delta T=8K$ for a free layer volume of $70 \times 70 \times 3 nm^3$), thermagnonic spin-transfer torque tends to align the free layer moment with $M_{Ferrite}$ asymptotically ($<m_z> \to 0$, *complete destabilization*). In Fig. 2(a-d), angular precession of $m_{Free}$ subject to a sub-critical thermal excitation ($\delta T=7K$), is compared with a super-critical case ($\delta T=10K$) which leads to complete free layer destabilization. The evolutions with time of $m_{Free}$ for both sub- and super-critical thermagnonic excitations are shown in Fig. 3(a) and 3(b) respectively. For $\delta T>8K$, the time required for complete free layer destabilization drops monotonically with increasing temperature differential as illustrated in Fig. 3(b). The sign of $\delta T$ has no effect on the delay associated with the free layer destabilization. With complete destabilization, the free layer moment either is aligned or anti-aligned with the reference layer moment, depending upon the direction of heat current flow. However, for low power and fast switching, partial destabilization is preferred, as will be discussed in subsequent sections.

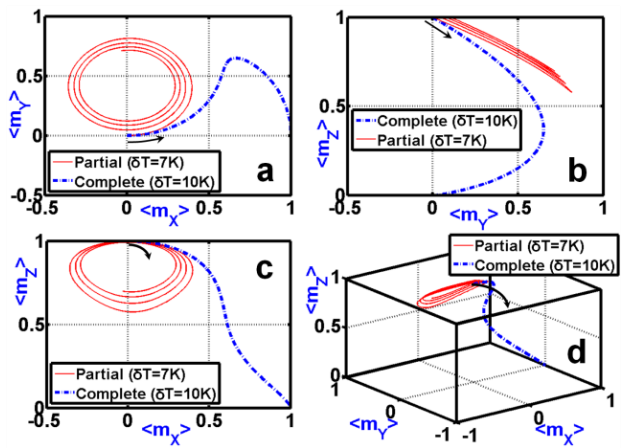

**Fig. 2** Angular evolution of the free layer moment, $m_{Free}$, in the (a) $m_X$-$m_Y$ plane; (b) $m_Y$-$m_Z$ plane; (c) $m_X$-$m_Z$ plane and (d) $m_X$-$m_Y$-$m_Z$ space for two different degrees of thermal excitation

Here, $H_{eff}$ is the total effective magnetic field acting on the free layer, $\gamma$ is the gyromagnetic ratio (17.6MHz/Oe) and $\alpha$ is the Gilbert damping constant which we set at 0.01. $H_{Ku2}$ and $H_{Ks}$ are the uniaxial (PMA) and easy-

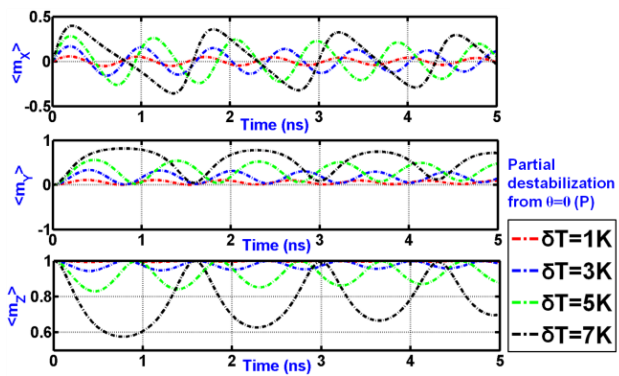

**Fig. 3a** Transient evolution of the free layer moment ($m_{Free}$) for sub-critical thermagnonic excitations

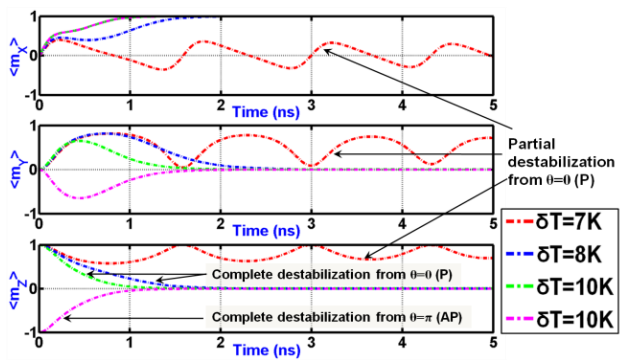

**Fig. 3b** Transient evolution of the free layer moment ($m_{Free}$) for super-critical thermagnonic excitations

## IV. Thermagnonic Spin-Transfer Torque MRAM: Electro-Thermal Design and Device Simulation

### A. Basic Structural Topology and Operating Principle:

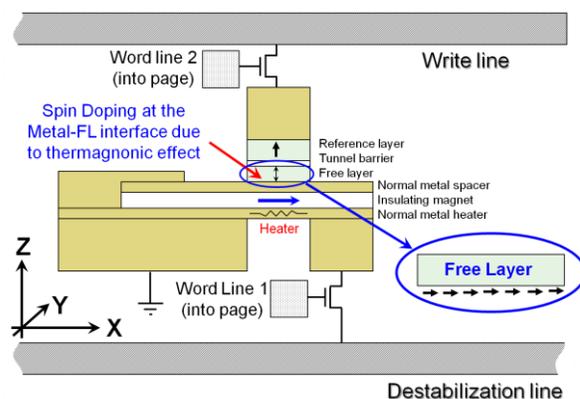

**Fig. 4** The basic structural topology of the proposed memory bit-cell utilizing magnonic spin-transfer torque for initial destabilization.

The basic structure of the proposed thermagnonic STT-MRAM is schematically shown in Fig. 4. The conventional tri-layer MTJ (reference-layer/tunnel-barrier/free-layer) with PMA is placed on top of a metal-ferrite junction which is in contact with a joule heater. To switch the free layer magnetic moment, bias is applied to the destabilization line and then word line 1 is turned on. A lateral electric current (y-direction) through the metallic heating element initiates a heat current that flows normal to the metal-ferrite interface (z-direction). Once the free layer moment is partially destabilized from its initial stable position, word line 1 is turned off, the bias is removed from the destabilization line, bias is applied to the write line, and then word line 2 is turned on, sending a small electric current through the multi-layer stack. This electrical current is low enough not to cause tunnel barrier breakdown and, since it is bipolar, can switch the free layer moment into either the parallel or anti-parallel state. The highly efficient thermagnonic torque overcomes the initial energy barrier associated with the free layer anisotropy. This reduces the required electric current density which flows through the tunnel barrier, enabling low power, fast, and almost error-free switching.

The proposed thermagnonic STT-MRAM has the Joule heater on top, for reasons which will be discussed, as shown in Fig. 5. The electric currents producing Joule heat and inducing spin-transfer torque are independently controlled by the transistors $T_{X,H}$ and $T_{X,E}$ respectively. The bias voltages required for reading and writing the proposed bit-cell are summarized in Table 1.

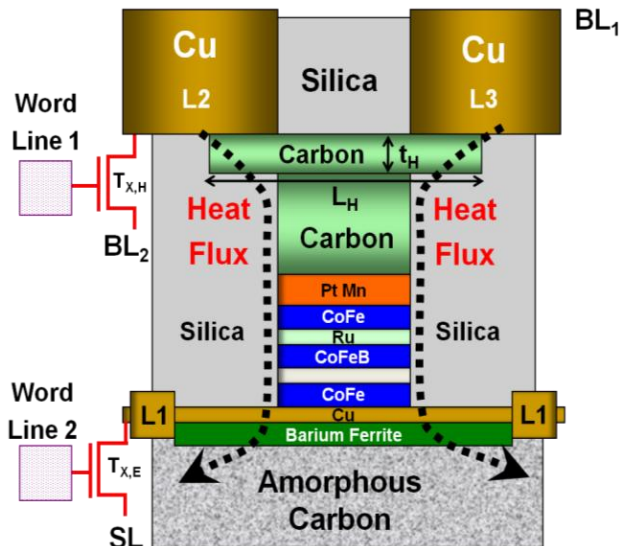

**Fig. 5** Magnonic MRAM structure. The joule-heater is on top. Electrically resistive carbon is used as the hard capping top layer (not to scale). The joule heating current between the copper electrodes ($L_2$, $L_3$) is electrically controlled by the transistor $T_{X,H}$. During data read and write, electrical current through the magnetic stack (reference layer/oxide/free layer) is controlled by the access device $T_{X,E}$. When $T_{X,H}$ and $T_{X,E}$ are turned ON and OFF respectively during the destabilization, the heat current funnels through the multi-layer stack yielding a temperature gradient at the Cu/Barium Ferrite interface.

Table 1: Voltage controlled read/write mechanisms in thermagnonic STT-MRAM

| Operation | | word line 1 | word line 2 | $BL_1$ | $BL_2$ | SL |
|---|---|---|---|---|---|---|
| **Write '0'** (AP-P) | Phase: 1 (destab.) | $V_{DD}$ | GND | GND | $V_{DD}$ | GND |
| | Phase: 2 | GND | $V_{DD}$ | GND | GND | $V_{Write}$ |
| **Write '1'** (P-AP) | Phase: 1 (destab.) | $V_{DD}$ | GND | GND | $V_{DD}$ | GND |
| | Phase: 2 | GND | $V_{DD}$ | $V_{Write}$ | GND | GND |
| **Read** | - | GND | $V_{DD}$ | $V_{Read}$ | GND | GND |

In the first phase of the write cycle (destabilization phase), word line 1 is turned on, keeping word line 2 off. A short duration (pulsed) voltage differential (~$V_{DD}$) between $BL_1$ and $BL_2$ develops a heating current through the resistive heating element made of carbon. The heat flux funnels through the magnetic multi-layers towards the sink at the bottom (lead $L_1$). Once a temperature gradient ($\delta T$) develops across the metal-ferrite interface which is sufficient for thermagnonic destabilization, word line 1 is turned off. Next, during the second phase, word line 2 is turned on just as, or even slightly before, word line 1 is turned off. With an appropriate voltage differential, $V_{Write}$, between $BL_1$ and SL, a small electric current density (J) flows through the magnetic tri-layer and spin-torque switches the free layer. For data read-out, word line 2 is turned on with $BL_1$ set to a voltage $V_{Read}$ and SL grounded. With a choice of higher resistance-area (RA) product for the tunnel barrier (MgO), it might be possible to make $V_{Read}$ equal to $V_{DD}$ [23]. This would eliminate the necessity of a second lower supply voltage as required in case of a conventional two terminal device [24]. The design approach with higher RA will be discussed in detail in section V.

B. *Impact of Thermagnonic Destabilization on the Switching Dynamics of the Free Layer Moment:*

The energy consumption of the proposed thermagnonic STT-MRAM is determined by the Joule heating from the heating current (~$J_H$) during the first phase of the write cycle, and the electrical spin torque current (~$J_E$) during the second phase of the write cycle. To estimate total dissipation we calculate the transient electro-thermal conduction coupled with the LLGS equation, assuming that the layers are single magnetic domains [10]. First, we solve the Fourier heat transfer equation [25] coupled with Ohm's and Joule's laws using a commercial finite element analysis (FEA) package [26]. In the analysis of the structure shown in Fig. 5, the material stack is assumed as follows: Amorphous Carbon(300nm)/BaFe2O4(30nm)/Cu(5nm)/CoFe(3nm)/MgO(1nm)/CoFeB(3nm)/Ru(1nm)/CoFe(2nm)/PtMn(15nm)/Carbon(50nm)/

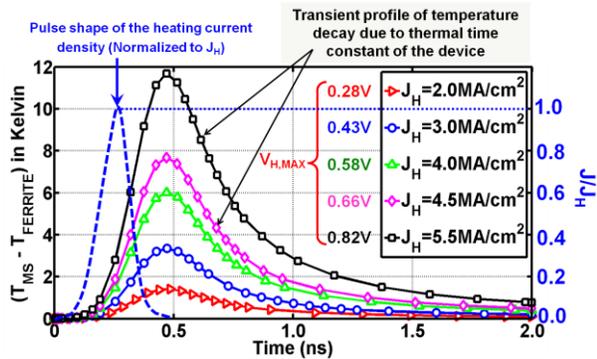

**Fig. 6** Response of the interface (Ferrite/Metal) temperature differential (δT) due to heating current (left-scale). The shape of the heating current pulse is assumed to be Gaussian with pulse width PW=0.5ns and amplitude $I_H = J_H * A_H$, where $A_H$ is the cross section of the carbon heating element (right-scale). The maximum voltage differential ($V_{H,MAX}$) between $BL_1$ and $BL_2$ has been indicated for different heating current densities ($J_H$).

Carbon($t_H$=30nm). The diameter of the circular MTJ pillar was assumed to be 70nm. After subjecting the device to a pulsed heating current (modeled as a Gaussian pulse of width PW=0.5ns, and of variable amplitude), Fig. 6 shows the response of the interfacial (metal-ferrite) temperature difference δT. For partial thermagnonic destabilization of the free layer moment, a heating current of sufficiently large amplitude but short pulse width is desired. Once the required temperature differential δT is developed at the metal-ferrite interface, the heating current should be turned off immediately, in order to conserve energy. The thermal time constant of the system delays the decay of δT, as shown in Fig. 6. After the heat pulse destabilizes the free layer moment from its initial uniaxial direction, δT starts dropping from its peak value. The injection of a small bi-polar electric current ($I_E$) through the magnetic tunnel junction completes the magnetization reversal process.

For efficient switching of the free layer, it is imperative to have a large magnon contribution to the heat flow across the metal-ferrite interface. In a very recent communication [28], the time-resolved heat-flow dynamics in ferromagnetic thin films have been studied experimentally. A large magnonic contribution to the total heat current flowing normal to the film surface has been reported over a broad temperature range. Subject to a pulsed heating, the local magnon temperature rises almost instantaneously due to fast (~ps) thermal energy transfer from the heat source.

Using the calculated transient profile for δT as input (Fig. 6), we solve the LLGS equation including both electrical and thermagnonic spin-transfer torque terms, as given in Fig. 7. The effect of thermally induced stochastic magnetic noise is modeled as in Brown [27]. To include the effect of thermal fluctuations and electrically induced spin-transfer torque [5-6] in addition to the thermagnonic contribution, we extend equation 4 as indicated in Fig. 7.

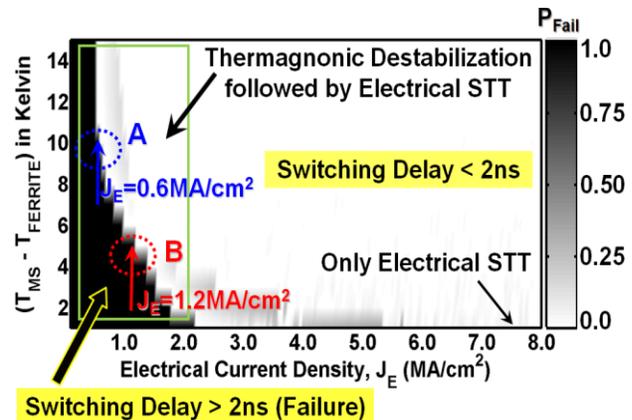

**Fig. 8** Free layer switching failure probability, $P_{Fail}$, in $J_E$-δT design space for a switching delay of 2ns. The black region indicates the failure of free layer switching (P-AP) within a write window of 2ns. Higher δT (~10K) at design point 'A' requires substantially lower $J_E$ than point 'B' with a relatively lower δT (~5K). Without magnonic destabilization (δT=0K), the critical switching current density ($J_E$) becomes unacceptably high (~8MA/cm$^2$) for a delay of 2ns associated with a switching failure probability as low as 1E-9.

Using the coupled equations described in Fig. 7, we estimate the joint impact of thermagnonic and electric spin-transfer torque on the free layer magnetic switching in Fig. 8. In a conventional STT-MRAM with PMA, the critical switching current density ($J_E$) for a delay of 2ns associated with a switching failure probability of 10$^{-9}$ is estimated to be close to 8MA/cm$^2$ at 300K. For an RA of 10 Ω-μm$^2$, this corresponds to a voltage across the MTJ of 800 mV, far more than the limit of 400 mV required to avoid breakdown over ten years [11, 30]. However, with thermagnonic destabilization, reducing the switching delay below 2ns with error-free switching and long term tunnel barrier reliability seems achievable. In Fig. 9, we analyze the switching failure characteristics of MRAM with thermagnonic destabilization, as a function of the electric current density ($J_E$). We compare the switching failure probability characteristics of the

$$\frac{d\vec{m}_{Free}}{dt} = -|\gamma|(\vec{m}_{Free} \times \vec{H}_{eff}) + \alpha(\vec{m}_{Free} \times \frac{d\vec{m}_{Free}}{dt})$$
$$+ a_J[-\alpha(\vec{m}_{Free} \times \vec{M}_{Ref}) + (\vec{m}_{Free} \times \vec{m}_{Free} \times \vec{M}_{Ref})] \leftarrow \text{Slonczewski Spin Torque Term [5]}$$
$$+ |\tau_{x0}|[-\alpha(\vec{m}_{Free} \times \vec{M}_{Ferrite}) + (\vec{m}_{Free} \times \vec{m}_{Free} \times \vec{M}_{Ferrite})] \leftarrow \text{Slonczewski Magnonic Torque Term [12]}$$

$$\vec{m}_{Free} = \frac{\vec{M}_{Free}}{M_S}; a_J = |\gamma| \cdot \frac{\hbar \cdot S_f \cdot J_E}{q \cdot M_S \cdot t_{FL}}; S_f = \left[ -4 + (1+\eta)^3 \frac{(3+\cos\theta)}{4\eta^{3/2}} \right]^{-1}$$

$$\vec{H}_{eff} = \vec{H}_{det} + \vec{h}_{thermalfluc}$$
$$\vec{H}_{det} = H_{x,External} \cdot \hat{x} + H_{y,External} \cdot \hat{y} + [H_{z,External} + H_{Ku2} - H_{Ks}] \cdot \hat{z}$$
$$\vec{h}_{thermalfluc} = h_{x,fluc} \cdot \hat{x} + h_{y,fluc} \cdot \hat{y} + h_{z,fluc} \cdot \hat{z}$$
$$h_{x,fluc}, h_{y,fluc}, h_{z,fluc} \sim Gaussian(\mu, \sigma)$$
$$\mu = 0; \sigma = dt^{-0.5} \sqrt{\frac{8\pi \cdot kT \cdot \alpha}{(1+\alpha^2) \cdot |\gamma| \cdot M_s \cdot V \cdot \zeta_{permeability}}}$$

Brown's Model [27] using Fluctuation-Dissipation Theorem and Langevin's Equation of Brownian Motion

**Fig. 7** Simulation framework for single domain magnetization dynamics using LLGS in the presence of thermally induced stochastic magnetic field [27] and thermagnonic spin-transfer torque.

thermagnonic STT-MRAM at design points 'A' (δT=10K) and 'B' (δT=5K) as marked in Fig. 8 with that of a conventional STT-MRAM with no thermagnonic excitation (δT=0K). The steep switching failure characteristic (blue and black lines in Fig. 9) is the most attractive feature of the proposed thermagnonic STT-MRAM. A marginal destabilization with an interfacial temperature differential (δT) of 5K helps in achieving almost 6X lower critical current density ($J_E$) than the conventional PMA device for a switching delay of 2ns, and with an associated switching failure probability lower than $10^{-9}$. The slope appears to be vertical, suggesting negligible switching failure probability is possible for a small overdrive in electrical current, consistent with the intuitive picture of deterministic destabilization followed by deterministic switching. This is in contrast to the inherently unreliable and non-deterministic switching used in conventional STT-MRAM, due to the absence of spin torque when the free and reference layers are collinear. Substantial reduction of critical switching current density ($J_E$~1.2MA/cm$^2$ @$T_{Write}$=2ns) allows the incorporation of a thicker tunnel barrier, enabling higher TMR [29] and long term tunnel barrier reliability.

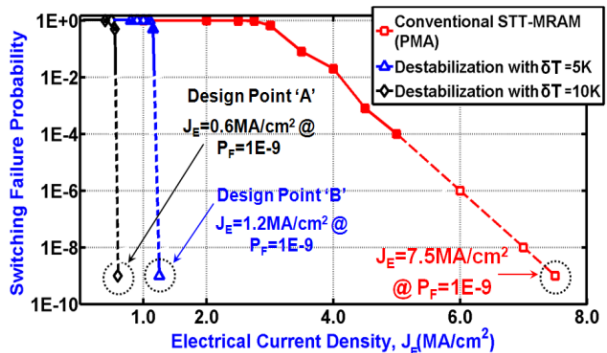

**Fig. 9** Comparison of switching failure characteristics of the proposed thermagnonic MRAM with that of conventional STT-MRAM with PMA. With ultra-steep failure probability characteristics (blue and black lines), the proposed MRAM outperforms the conventional STT device with almost 7X lower switching current density and higher tunnel barrier reliability.

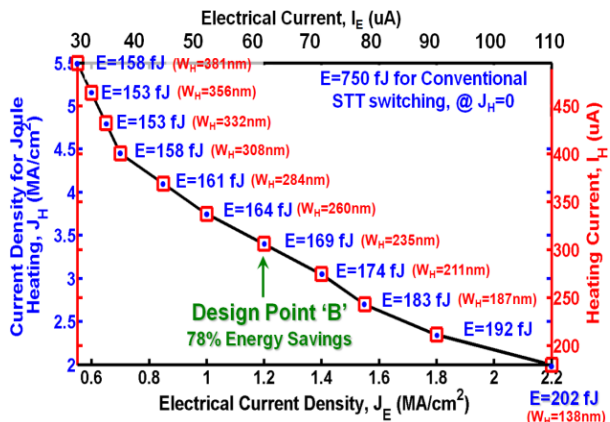

**Fig. 10** Design optimization space ($J_H$ vs. $J_E$) of the proposed thermagnonic MRAM for a switching delay of 2ns and a switching failure probability of $10^{-9}$. The simulations assume a 32nm process [34], ($V_{HEATING,MAX}$)=1.2V; ($V_{E,MAX}$)=0.5V, $E_a$=85kT, $M_{Sat}$=850emu/cm$^3$, $H_{Ku2}$=880emu/cm$^3$, vol.=70x70x3nm$^3$.

In Fig. 10, we estimate the total switching energy E of the proposed device for discrete design points with different heating ($J_H$) and electric current densities ($J_E$). Assuming a 32nm process for the heating device ($T_{X,H}$ in Fig. 5) and $V_{DD}$=1.0V, the required transistor widths ($W_H$) are indicated in Fig. 10 for each design point. Total switching energy E (including dissipation from both $J_E$ and $J_H$) gradually increases with reduced thermagnonic contribution for constant delay (2ns) and switching failure probability ($10^{-9}$). For a design point close to 'B' ($\delta T_C$=5K, $J_E$=1.2MA/cm$^2$), $W_H$ can be kept around 250nm with a total energy savings of 78% relative to the conventional case with no thermagnonic destabilization. Further gains in energy savings would require prohibitively large transistors to drive the heater element. For higher memory integration density, $W_H$ can be scaled to smaller dimensions (~140nm) with a marginal (~5%) sacrifice in energy savings.

## V. Read Optimization in Thermagnonic Spin-Transfer Torque MRAM

The presence of an access transistor ($T_{X,E}$ in Fig. 5) in series with the MTJ during read reduces the effective TMR of the bit-cell, which can be expressed as:

$$\text{TMR}_{\text{CELL}} = \frac{\text{TMR}_{\text{MTJ}}}{(1 + \frac{R_{\text{NMOS}}}{R_P})} \quad (5.a)$$

$$\text{TMR}_{\text{MTJ}} = \frac{R_{AP} - R_P}{R_P} \quad (5.b)$$

$R_{NMOS}$, $R_{AP}$ and $R_P$ are the resistance of the access device during read, and anti-parallel and parallel resistances of the MTJ respectively. To ensure a high cell TMR during read, one needs to maximize both $R_P$ and $TMR_{MTJ}$ for a fixed width of the access transistor [31]. This can be done by increasing the thickness of the tunnel barrier in the MTJ pillar.

Fig. 11(a-d) shows the effect of tunnel barrier thickness ($T_{OX}$) and voltage on $R_{AP}$, $R_P$, TMR and tunneling current density (J) of a standard tri-layer MTJ with MgO as tunnel barrier. The details of spin-dependent electronic transport simulation using Non-Equilibrium Green's Function (NEGF) formalism [32] have been reported in [33]. As shown in Figs. 11(a) and (b), the differential resistances (dV/dI) in parallel and anti-parallel states of an MTJ increase by almost an order of magnitude when $T_{OX}$ is changed from 1.2nm to 1.5nm. The increase in tunnel barrier thickness also helps in boosting the device TMR, as illustrated in Fig. 11(c). Thicker tunnel barriers are not feasible in conventional STT-MRAM, given a switching current density requirement of order 8MA/cm$^2$. However, RA

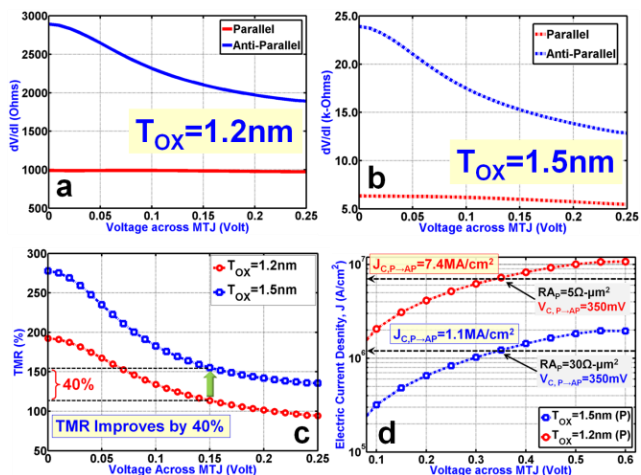

**Fig. 11** Bias-voltage dependence of parallel and anti-parallel differential resistances (dV/dI) of an MTJ with (a) $T_{OX}$=1.2nm and (b) $T_{OX}$=1.5nm. (c) A thicker tunnel barrier helps in boosting up the MTJ TMR. (d) Bias-voltage dependence of electric current density (J) through the tunnel barrier for two different oxide thicknesses.

torque for bi-directional magnetic switching. The combination of lower power, improved reliability, higher integration density, and larger read margin makes the proposed MRAM a competitive choice for future non-volatile memory technology. As compared to the conventional STT-MRAM with PMA, thermagnonic STT-MRAM reduces the magnetization switching energy by almost 80% for a switching delay of 2 ns and a switching failure probability less than $10^{-9}$.

**Acknowledgement:** This research was supported in part by an IBM Ph.D fellowship and the Nano-electronics Research Initiative (NRI) under the INDEX center. The authors would like to sincerely thank John Slonczewski for his helpful technical suggestions.

products can be increased in thermagnonic STT-MRAM due to the substantially lower critical switching current density (~1.2MA/cm$^2$) for the same delay and switching probability. With a marginal increase in MgO thickness ($T_{OX}$~1.5nm), RA products in P and AP states can be raised to 30 Ω-μm$^2$ and 70 Ω-μm$^2$ respectively, still sufficient for fast (~2ns) and error-free magnetic switching. The thicker tunnel barrier provides an enhanced guard band against 'soft' oxide breakdown [30] and also helps in achieving almost 40% higher MTJ TMR [Fig. 11(c)].

Table 2: Thermagnonic MRAM vs. conventional PMA STT-MRAM

| Memory design attributes for a free layer vol.=70x70x3nm$^3$ | Thermagnonic MRAM Bit-cell: ($T_{OX}$=1.5nm) | STT-MRAM Bit-cell: ($T_{OX}$=1.2nm) |
|---|---|---|
| STT switching delay (Parallel →Anti-Parallel) | 2.0 ns | 2.0 ns |
| Electrical switching current density ($J_E$)/ current ($I_E$) @2ns | $J_E$=1.2MA/cm$^2$ $I_E$ =50μA | $J_E$=7.5MA/cm$^2$ $I_E$ =350μA |
| Heating current density ($J_H$)/ current ($I_H$) @2ns | $J_H$=3.75MA/cm$^2$ $I_H$ =350μA ($\delta T_C$=5K) | - |
| Total switching energy/bit ($J_T$) @$T_{WRITE}$ = 2ns and $P_{Fail}$=10$^{-9}$ | $J_E$=60 fJ $J_H$=109 fJ $J_T$= $J_E$+$J_H$=169 fJ | $J_E$=750 fJ $J_H$=0 $J_T$= $J_E$=750 fJ |
| MTJ TMR @ $V_{MTJ}$=0.1V | 155% | 115% |
| Resistance-Area (RA) in P and AP configuration | $RA_P$=30Ω-μm$^2$ $RA_{AP}$=70Ω-μm$^2$ | $RA_P$=5Ω-μm$^2$ $RA_{AP}$=10Ω-μm$^2$ |

In summary, we show a detailed comparison of the proposed thermagnonic STT-MRAM bit-cell with that of conventional STT-MRAM in Table 2.

### VI. Conclusions

In this paper, we proposed the design of a new genre of STT-MRAM utilizing thermagnonic spin-transfer